\documentclass[fleqn,usenatbib]{mnras}

\usepackage{newtxtext,newtxmath}

\usepackage[T1]{fontenc}
\usepackage{ulem}

\DeclareRobustCommand{\VAN}[3]{#2}
\let\VANthebibliography\thebibliography
\def\thebibliography{\DeclareRobustCommand{\VAN}[3]{##3}\VANthebibliography}

\usepackage{graphicx}	
\usepackage{amsmath}	

\title[Spin and orbital angular momentum]{An apparent positive relation between spin and orbital angular momentum in X-ray binaries}

\author[Z. Yan et al.]{
Zhen Yan,$^{1}$\thanks{E-mail: zyan@shao.ac.cn}
Wenda Zhang,$^{2}$
Wenfei Yu$^{1}$
\\
$^{1}$Shanghai Astronomical Observatory, Chinese Academy of Sciences, Shanghai 200030, China  \\
$^{2}$ Key Laboratory of Space Astronomy and Technology, National Astronomical Observatories, Chinese Academy of Sciences, Beijing 100012, China
}

\date{Accepted XXX. Received YYY; in original form ZZZ}

\pubyear{2022}

\begin{document}
\label{firstpage}
\pagerange{\pageref{firstpage}--\pageref{lastpage}}
\maketitle

\begin{abstract}
The origin of current angular momentum (AM) of the black hole (BH) in X-ray binary (XRB) is still unclear, which is related with the birth and/or the growth of the BH. Here we collect the spin parameters $a_{*}$ measured in BH XRBs and find an apparent bimodal distribution centered at $\sim$ 0.17 and 0.83. We find a positive relation between the spin parameter and the orbital period/orbital separation through combining distinct XRB categories, including neutron star (NS) low-mass X-ray binaries (LMXBs), Roche-lobe overflow (RLOF) BH XRBs and wind-fed BH XRBs. It seems that the AM of the compact star and the binary orbit correlates by combining the different XRB systems. These positive relations imply that accretion process is a common mechanism for spinning up the compact star in these diverse XRB systems. We infer that the low and high spin BH XRBs may experience different evolution and accretion history, which corresponds to the bimodal distribution of the BH spin parameters. The low spin BHs ($a_{*}<0.3$) are similar to the NS LMXBs, the compact star of which is spun-up by the low-level accretion, and the high spin BHs ($a_{*}>0.5$) had experienced a short hypercritical accretion ($\gg \dot{M}_\mathrm{Edd}$) period, during which, the BH spin dramatically increased. 
\end{abstract}




\section{Introduction} \label{sec:intro}

 The dimensionless spin parameter $a_{*}$ is one of three physical parameters of a black hole (BH). The current spin parameter reflects the formation and evolution history of the BH.
The stellar-mass BH is generally believed to form from a core collapse of a massive star \citep[e.g.][]{woosley_gamma-ray_1993,oconnor_black_2011}. 
However the spin of the newborn BH is still indeterminable, since there are still many unresolved issues about the detailed formation processes \citep{Tauris_2006,Han_2020}, for example what fraction of mass and angular momentum (AM) are lost during the collapsing. If the current BH spin is natal, the relation between the spin and the parameters of the binary systems must give us some clues of the the properties of the progenitor \citep[e.g.][]{lee_discovery_2002, moreno_mendez_kerr_2011}. On the other hand, it has been well known that the BHs in X-ray binaries can be spun up by accretion \citep{bardeen_kerr_1970,thorne_disk-accretion_1974}. The gas from the companion star forms an accretion disc, which will lead to a strong and stable accretion torque acting on the central BH. The final spin will be determined by the total accreted mass \citep{thorne_disk-accretion_1974,belczynski_black_2008}. The mass transfer also relates with the binary parameters, such as the orbital period, the companion mass and radius. Whether the current BH spin reflects the accretion history or natal spin, however, is still under debate. 
 
The number of published stellar-mass BH spin parameter measurements has exploded in the past two decades, especially from the merging of binary black holes \citep[BBH; ][]{collaboration_gwtc-3_2021}. 
Before that all the measurements of stellar-mass BH spin parameter are from BH X-ray binaries (XRBs). In BH XRBs the stellar-mass BH accretes material from its non-degenerate companion and form an accretion disc around the BH. So far two main methods have been proposed to measure the BH spin parameter based on the accretion process. Both methods assumed that the accretion disc extends down to the innermost stable circular orbit (ISCO), the radius of which ($R_\mathrm{ISCO}$) monotonically decreases with the increasing BH spin. One is the continuum fitting method, which was first proposed by \citet{zhang_black_1997}. The basic idea is to fit the thermal X-ray emission from the thin accretion disks to determine the $R_\mathrm{ISCO}$ \citep{mcclintock_black_2014}. The other is the relativistic reflection method, which is to determine the $R_\mathrm{ISCO}$ according to the general relativistic effect through modeling the asymmetrical iron $K\alpha$ line and the reflection component \citep{fabian_x-ray_1989,reynolds_measuring_2014}. 
There is another X-ray timing method, which was developed by \citet{motta_black_2014,motta_precise_2014}, has only been applied to few systems. Since this method needs the rare detection of high-frequency qusi-periodic oscillations (QPOs)  \citep{remillard_x-ray_2006,belloni_fast_2014} simultaneous to the low-frequency QPOs. Recently, \citet{bhargava_timing-based_2021} has shown this technique has a broad application prospect by using the high-frequency broad band noise instead of high-frequency QPO.

The relations between the BH spin and binary parameters (e.g. orbital period, masses) will provide valuable information about the BH formation process and/or the mass transfer history. 
Previous studies have shown that many properties of the BH XRBs are related with the orbital period. For example, \citet{lee_discovery_2002} has found a correlation between the BH mass and orbital period and used to predict the BH spin. \citet{gandhi_astrometric_2020} has found that the Galactic height is anti-correlated with the binary period for known BH XRBs. 
\citet{fragos_origin_2015} has shown a positive relation between the BH spin and orbital period, and they thought this relation is consistent with the accretion spinning up phenomenon. However, they only used the BH spin measurements from the continuum fitting method. In this work, we are going to investigate the relations between the BH spin and binary parameters in BH XRBs by using the latest BH spin measurements. We also include a sample of neutron star low mass X-ray binaries (NS LMXBs), since the NS of which is generally thought to be spun up by the accretion \citep[see reviews in][]{srinivasan_recycled_2010}. There probably are some similarities between NS and BH accreting systems. The comparison between these two will highlight the origin of the spin.

\section{Source sample and binary system parameters}\label{sys_par}
\begin{figure}
\includegraphics[width=\linewidth]{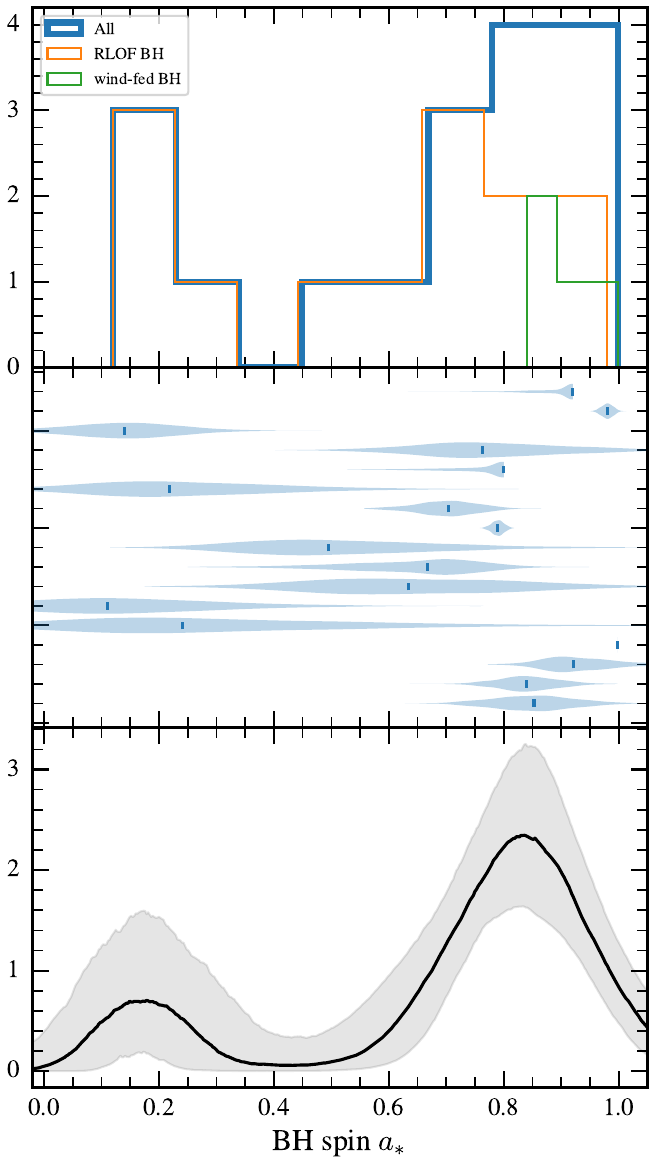}
\caption{Top: the distribution of the measured spin parameters $a_{*}$ of BH XRBs. Middle: resampling the data by considering the errors and upper/lower limits. Bottom: the double Gaussian distribution and the 1$\sigma$ range.   }
\label{fig:hist}
\end{figure}

\begin{table*}
 \caption{ Binary system parameters of BH XRBs}
 \label{tab:bh_par}
 \begin{tabular}{lcccccc}
  \hline
  \hline
  Source & $a_{*}$ & $P_\mathrm{orb}$ & $M_\mathrm{1}$ & $q$  &Ref. \\
          &        &   (days)  & ($M_{\sun}$) & ($M_{2}/M_{1}$)  & \\
  \hline
  \hline
IC 10 X-1 & $0.85^{+0.04}_{-0.07}$ &1.455  &32.7$\pm$2.6 &1.07$\pm$0.08  & [1][2] \\
M33 X-7  & $0.84\pm 0.05$  & 3.453    & $15.65\pm1.45$    &$4.47\pm0.61$       &[3][4][5][6] \\
Cyg X-1  &  $>0.9985$ (0.001)$^{*}$  &5.600 &$ 21.2 \pm 2.2$  &   $1.92\pm 0.41$    &[7][8][9]  \\   
LMC X-1 & $0.92^{+0.05}_{ -0.07}$   & 3.909  &$10.91\pm 1.41$   &$2.92\pm0.50$      &[10][11][12][13]    \\
NGC300 X-1 & ... & 1.366 & 17.0$\pm$4.0. & $1.53\pm0.55$ &  [14]\\
\hline
LMC X-3 & $0.25^{+0.20}_{-0.29}$    &1.705  &$ 6.98 \pm 0.56$  &$0.52\pm0.09$       &[15][16]\\
A 0620$-$00         & $0.12\pm 0.19$   &0.323   &$5.86\pm0.24$   &$0.06\pm0.007$       & [17][18][19] \\
GS 1124$-$683    &$0.63^{+0.16}_{-0.19}$    &  0.433  &$11.0^{+2.1}_{-1.4}$ &$0.079\pm0.007$           & [20][21][22] \\
4U 1543$-$47   & $0.67^{+0.15}_{-0.08}$   & 1.123  &$5.1\pm2.4$  &$0.48\pm0.23$           &[23][24]   \\
XTE J1550$-$654  & $0.49^{+0.13}_{-0.2}$  &1.542 &$9.10\pm0.61$ &$0.034\pm0.007$      &[25][26] \\
XTE J1650$-$500    &$0.79\pm0.01$     &0.321      & 5.65$\pm$1.65    &0.1      & [27] [28] \\
GRO J1655$-$40  & $0.7\pm0.05$    & 2.622 &$5.99\pm0.42$   &$0.419 \pm 0.028$       &  [29][30][31]  \\
MAXI J1659$-$152 &$0.21^{+0.14}_{-0.2}$ &0.101 &5.4$\pm$2.1&0.045$\pm$0.025 &  [32][33]\\
GX 339$-$4          &  $\lesssim0.8$ (0.1)$^{*}$ & 1.759    &7.5$\pm$3.5              & 0.13$\pm$0.08      &[34][35] \\ 
Swift J1753.5$-$0127    & $0.76^{+0.11}_{-0.15}$ &0.119   &$\gtrsim 7.4\pm1.2$   &$\gtrsim0.04\pm0.03$    &[36][37][38]  \\
MAXI J1820+070 & 0.14$\pm0.09$    &0.685   &$8.48^{+0.79}_{-0.72}$   &0.072$\pm$0.012  &    [39][40][41] \\
GRS 1915+105  &  $0.98\pm0.01$  &33.850 &$12.4^{+ 2.0}_{-1.8}$  &   $0.042 \pm 0.024$     &[42][43] \\   
V404 Cyg & $>0.92$ (0.078)$^{*}$ &6.471 & $9.0^{+0.2}_{-0.6}$ & $0.06\pm 0.005$ &[44][45] \\
GRO J0422+32 & ... & 0.212  & 4.0$\pm$1.0  & 0.0433  &    [46]\\
GS 1009$-$45 & ... & 0.283 &4.2$\pm$0.6 &0.137$\pm$0.015   &   [47]\\
XTE J1118+480 & ... & 0.170 & 7.55$\pm$0.65 & 0.0137$\pm$0.001 &    [48] \\
MAXI J1305$-$704 & ... & 0.394  & 8.9$\pm$1.5 & 0.048$\pm$0.02 &  [49]\\ 
GS 1354$-$64 & ... & 2.544 & 7.6$\pm$0.7 & 0.12$\pm$0.04   &  [50]\\
Swift J1357.2$-$0933 & ... & 0.110 & 12.4$\pm$3.6 &0.04$\pm$0.02 &     [51] \\
H 1705$-$250 & ... & 0.521 & 6.4$\pm$1.5 & 0.053 &     [52][54]\\
XTE J1859$+$226 & ... & 0.274 & 7.7$\pm$1.3 & 0.09  &   [53]\\
GS 2000+251 & ... & 0.344 & 7.2$\pm$1.7 & 0.04$\pm$0.1  &    [54]\\
V4641 Sgr & ... & 2.817 &6.4$\pm$0.4 & 0.67$\pm$0.04 &     [55]\\
  \hline
  \hline
 \end{tabular}
 \\

Note: The spin parameter uncertainties of Cyg X-1, GX 339-4 and V404 Cyg are assumed as the values in the brackets.
\\
{Ref. }  
[1]\citet{steiner_spin_2016};[2]\citet{silverman_ic_2008};
[3]\citet{liu_precise_2008}; [4]\citet{liu_erratum:_2010}; [5]\citet{pietsch_m33_2006}; [6]\citet{orosz_15.65-solar-mass_2007};
[7]\citet{miller-jones_cygnus_2021}; [8]\citet{zhao_re-estimating_2021};  [9]\citet{brocksopp_improved_1999}; [10]\citet{gou_determination_2009}; [11]\citet{steiner_broad_2012}; [12]\citet{levine_detection_2006}; [13]\citet{orosz_new_2009};[14]\citet{binder_wolf-rayet_2021};[15]\citet{steiner_low-spin_2014}; [16]\citet{orosz_mass_2014};  [17]\citet{gou_spin_2010}; [18]\citet{gonzalez_hernandez_doppler_2010}; [19]\citet{van_grunsven_mass_2017}; [20]\citet{chen_spin_2016}; [21]\citet{orosz_improved_1996}; [22]\citet{wu_mass_2016}; [23]\citet{dong_spin_2020}; [24]\citet{orosz_orbital_1998};  [25]\citet{steiner_spin_2011}; [26]\citet{orosz_improved_2011}; [27]\citet{miller_stellar-mass_2009}; [28]\citet{orosz_orbital_2004};  [29]\citet{shafee_estimating_2006}; [30]\citet{van_der_hooft_quiescence_1998}; [31]\citet{shahbaz_determining_2003}; 
[32]\citet{feng_estimating_2022}; [33]\citet{torres_delimiting_2021};
[34]\citet{zdziarski_x-ray_2019}; [35]\citet{kolehmainen_limits_2010};
  [36]\citet{reis_determining_2009}; [37]\citet{neustroev_spectroscopic_2014};
  [38]\citet{shaw_no_2016};
  [39]\citet{torres_dynamical_2019};
[40]\citet{torres_binary_2020};
[41]\citet{zhao_estimating_2021};
  [42]\citet{reid_parallax_2014};
  [43]\citet{steeghs_not-so-massive_2013};
[44]\citet{walton_living_2017};
[45]\citet{khargharia_near-infrared_2010};
[46]\citet{gelino_gro_2003}; [47]\citet{filippenko_black_1999}; [48]\citet{khargharia_mass_2013}; [49]\citet{mata_sanchez_dynamical_2021};[50]\citet{casares_refined_2009}; [51]\citet{casares_mass_2016}; [52]\citet{filippenko_black_1997}; [53]\citet{corral-santana_evidence_2011}; [54]\citet{casares_mass_2014};[55]\citet{macdonald_black_2014}  

\end{table*}

\begin{table*}
 \caption{Binary system parameters of NS LMXBs }
 \label{tab:ns_par}
 \begin{tabular}{lccccccc}
\hline
\hline
Source & $\nu_\mathrm{spin}$ & $P_\mathrm{orb}$ & $M_\mathrm{1}$ & $R_{1}$ &$M_{2}$   &Ref. \\
          &  (Hz)       &   (days)  & ($M_{\sun}$) & (km) & ($M_{\sun}$)   &                \\
\hline
\hline
EXO 0748$-$676      & 552  &0.1593  & $1.65\pm0.11$&$10.0\pm2.0$  &$0.45\pm0.17$ &[1][2][3]\\
4U 1608$-$522      & 620  &0.5370 &$1.57 \pm 0.30$  &$10.36 \pm 1.98$  &0.32$\pm$0.18       & [4][5][6]        \\ 
4U 1636$-536$ & 582 & 0.1580 & ... & ... & 0.47$\pm$0.17 & [7][8] \\
SAX J1748.9$-$2021 & 442 & 0.3654 &$1.81\pm0.37$ &$11.25\pm 1.78$  &$0.93\pm0.18$     &[9][10][11][6]      \\
XTE J1807$-$294  & 190  & 0.0279 &$1.60\pm0.50$ &13.0$\pm$3.0   &0.01$\pm$0.001   &[12][13][14]  \\
SAX J1808.4$-$3658 & 401 &  0.0839 & $1.50\pm0.60$ &$11.8\pm1.3$   &$0.09\pm0.02$    &[15][16][17][18][19]\\
XTE J1814$-$338  & 314  & 0.1779 &$1.90 \pm 0.70$ &$15.0\pm3.0$    &$0.35 \pm 0.05$      &[20][21][22][23]\\
IGR J00291+5934  & 599  & 0.1024  &...   &...     &0.039$^{*}$       &[24] \\
MAXI J0911$-$655   & 340  & 0.0308 & ... & ... &0.024$^{*}$      & [25]   \\
XTE J0929$-$314  & 185  & 0.0304  &...   &...    &0.0083$^{*}$     & [26]  \\
PSR J1023+0038 & 592 & 0.1979 & 1.71$\pm$0.16 & ... &0.24$\pm$0.02    &  [27][28]\\
XSS J12270$-$4859 & 593 & 0.2879 & 1.62$\pm$0.76 & ... & 0.31$\pm$0.15  & [29]\\
IGR J17379$-$3747 &468 & 0.0783 &... &... & 0.056$^{*}$   &  [30] \\
NGC6440 X-2 & 206 & 0.0396 &...   &...   &0.007$^{*}$      & [31]          \\
Swift J1749.4$-$2807 & 518 & 0.3675 &1.5$\pm$0.7    &...   &$0.64\pm0.17$  &[32]           \\
IGR J17498$-$2921 & 401 & 0.1601 &...   &...   &0.17$^{*}$        &[33]           \\
IGR J17511$-$3057 & 245 &  0.1446 &...   &...     &0.136$^{*}$      &[34][35] \\
XTE J1751$-$305 & 435 &  0.0295  &...  &...     &0.014$^{*}$     &[36]\\
Swift J1756.9$-$2508  & 182  &  0.0379 &...    &...    &0.007$^{*}$     &[37]\\
IGR J17591$-$2342 & 527 & 0.3667 & ... & ... & 0.359$^{*}$  & [38] \\ 
IGR J17062$-$6143 &164 & 0.0263 & ... & ... & 0.0056$^{*}$   & [39] \\
IGR J16597$-$3704 & 105 & 0.0321 & ... & ... & 0.0062$^{*}$ & [40] \\
IGR J18245$-$2452 & 254 & 0.4596 &...   &...   &0.179$^{*}$    &[41]\\
HETE J1900.1$-$2455  & 377  &0.0579 &...    &...  &0.016$^{*}$        &[42]\\
Aql X-1     & 550 & 0.7895 &$1.8\pm0.4$    &$10.2\pm1.5$  &0.74$\pm$0.22         &[43][44][45][3]                    \\ 
MAXI J1957+032&    314     &0.0423  &... &... &0.0147$^{*}$ & [46] \\
MXB 1659$-$298 &  567 & 0.2966 & $1.5\pm0.5$ &$13.0\pm7.0$ & 0.75$\pm$0.45 &   [47][48][3]\\
4U 1916$-$053  &  270  & 0.0346 & ... & ... & 0.09$\pm$0.01 &   [49][50]\\
MAXI J1816$-$195 & 528  &0.2014 & ... & ... & 0.1$^{*}$ &   [51]\\
  \hline
  \hline
 \end{tabular}
 \\
{Note:} The $M_{2}$ marked with $*$ is the minimum companion mass.
 \\
{Ref.:}  [1]\citet{galloway_discovery_2010}; [2]\citet{knight_eclipse_2022}; [3]\citet{marino_obtaining_2018};[4]\citet{hartman_discovery_2003}; [5]\citet{wachter_closer_2002}; [6]\citet{ozel_dense_2016};
[7]\citet{casares_detection_2006};[8]\citet{strohmayer_evidence_2002};
[9]\citet{altamirano_intermittent_2008}; 
[10]\citet{patruno_phase-coherent_2009}; [11]\citet{guver_mass_2013};
[12]\citet{kirsch_studies_2004}; [13]\citet{chou_precise_2008}; [14]\citet{leahy_constraints_2011};
[15]\citet{chakrabarty_two-hour_1998}; [16]\citet{wijnands_millisecond_1998}; [17]\citet{hartman_long-term_2008}; [18]\citet{goodwin_bayesian_2019}; [19]\citet{deloye_optical_2008};[20]\citet{markwardt_xte_2003}; [21]\citet{papitto_timing_2007}; [22]\citet{leahy_constraints_2009}; 
[23]\citet{baglio_long-term_2013};[24]\citet{galloway_discovery_2005}; [25]\citet{sanna_discovery_2017}; [26]\citet{galloway_discovery_2002};[27]\citet{archibald_radio_2009};[28]\citet{deller_parallax_2012};
[29]\citet{roy_discovery_2015}; [30]\citet{sanna_discovery_2018};[31]\citet{altamirano_discovery_2010}; [32]\citet{altamirano_discovery_2011}; [33]\citet{papitto_discovery_2011}; [34]\citet{markwardt_rxte_2009}; [35]\citet{riggio_timing_2011}; [36]\citet{markwardt_discovery_2002};[37]\citet{krimm_discovery_2007}; [38]\citet{sanna_nustar_2018};[39]\citet{strohmayer_nicer_2018};[40]\citet{sanna_discovery_2018};    [41]\citet{papitto_swings_2013}; [42]\citet{kaaret_discovery_2006}; [43]\citet{zhang_spectral_1998}; [44]\citet{chevalier_discovery_1991};
[45]\citet{mata_donor_2017}; [46]\citet{sanna_new_2022};
[47]\citet{wijnands_disco_2001}; [48]\citet{iaria_possi_2018};
[49]\citet{galloway_discovery_2001}; [50]\citet{iaria_signature_2015};
[51]\citet{bult_disco_2022}
\end{table*}

 The measured parameters of the BH XRBs is mainly collected from \citet{corral-santana_blackcat:_2016, jonker_observed_2021,reynolds_observational_2021} and the reference therein. Our sample also includes the most recent measurements, such as MAXI J1820+070 measured by $Insight$-HXMT \citep{zhao_estimating_2021}. In our BH XRB sample, there are four of them have more massive companion stars than the BHs (mass ratio $q=M_{2}/M_{1}>1$, where $M_{1}$ is the mass of the compact object, $M_{2}$ is the companion mass), the BH of which are fed by the strong stellar wind of the companion. They are also called BH high-mass X-ray binaries (HMXBs). Others have less massive companion stars than the BHs ($q<1$), which are Roche-lobe overflow (RLOF) systems. We use the most recent measurements of spin parameters, if the results from different methods are consistent within uncertainties.
However, the spin parameters from different methods are inconsistent in GRO J1655$-$40 \citep {motta_precise_2014}, GX 339$-$4 \citep{parker_nustar_2016,zdziarski_x-ray_2019}, MAXI J1820+070 \citep{bhargava_timing-based_2021}. We adopt the measurements from the continuum fitting method, since the free parameters are less in this method. The uncertainties of the BH spin of Cyg X-1 and V404 Cyg are assumed as 0.001 and 0.078 due to the theoretical upper limit of BH spin. The BH spin of GX 339$-$4 is given as $<$ 0.8 in \citet{zdziarski_x-ray_2019}, $<$0.9 in \citet{kolehmainen_limits_2010} and 0.935 in \citet{reis_systematic_2008}. So we take the most recent measurement 0.8 as the BH spin and assume the uncertainty as 0.1. For simplicity, the lower limits of the BH mass and mass ratio of Swift J1753.5-0127 are used as measured values in the following analysis.

The distribution of the spin parameters of BH XRBs are shown in \autoref{fig:hist}. The two kinds of BH XRBs (RLOF and wind-fed) show an apparent dichotomy in BH spin (see \autoref{tab:bh_par}). It is worth noting that current known spin parameters in wind-fed BH XRBs are all very high ($>0.8$), and the spin parameters of RLOF BH XRBs span a large range ($\sim0.1$--0.98,\autoref{tab:bh_par}). 
We then used the Bayesian inference package \texttt{UltraNest} \citep{buchner_ultranest_2021} to explore the intrinsic distribution of BH spin parameters including the asymmetric error bars and upper/lower limits. We consider three different distributions: uniform, Gaussian and double Gaussian. The marginal likelihood ($\ln Z$) is $-4.11\pm$0.07, $-6.22\pm$0.06 and $-1.8\pm$ 0.13, respectively. The Bayes factor between two models was calculated for model comparison $ B_{ij}=Z_{i}/Z_{j}$, where the $Z_{i}$ and $Z_{j}$ is the marginal likelihood of model $i$ and $j$, and $B_{ij}>10$ is a strong evidence for supporting model $i$ \citep{Jeffreys_1939}. So the Bayes factors ($\sim 10$ and $\sim 83$) strongly support the double Gaussian distribution, $r_{1}*Norm(\mu_{1},\sigma_{1})+(1-r_{1})*Norm(\mu_{2},\sigma_{2})$. The best-fitting parameters are $r_{1}=0.24^{+0.12}_{-0.10}$ $\mu_{1}=0.17^{+0.08}_{-0.08}$,$\sigma_{1}=0.07^{+0.05}_{-0.04}$, $\mu_{2}=0.83^{+0.04}_{-0.05}$ and $\sigma_{2}=0.12^{+0.05}_{-0.03}$  (see also the bottom panel of \autoref{fig:hist}).

Our NS LMXB sample only includes the systems containing the fast rotating NSs (spin frequency $>100$ Hz, \autoref{tab:ns_par}), which is identified either from pulsation or burst oscillation, are mainly collected from \citet{campana_accreting_2018,he_formation_2019,patruno_accreting_2021} and references therein. We only collected the sources with both spin frequency and orbital period measurements. Most sources in the NS LMXB sample are also referred as accreting millisecond X-ray pulsars (AMXPs). 


\section{Positive relations between spin and orbital parameters}
\label{sec:corr}
\begin{figure*}
\includegraphics[width=\linewidth]{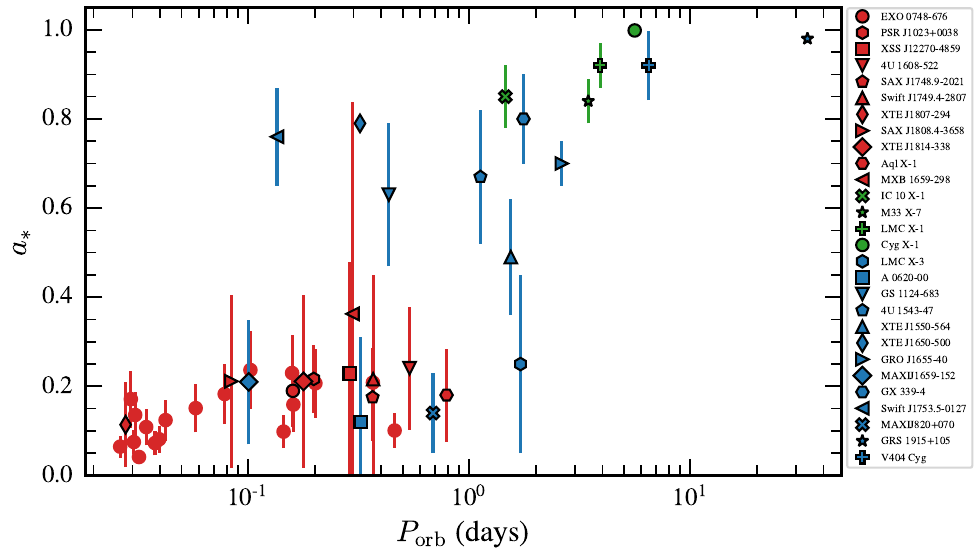}
\caption{The relation between spin parameter and orbital period of our samples. The NS, wind-fed BH and RLOF BH systems are marked by red, green and blue, respectively. The red dots are the NS LMXBs without mass and/or radius measurements.
\label{fig:porb}}
\end{figure*}

\begin{figure*}
\includegraphics[width=\linewidth]{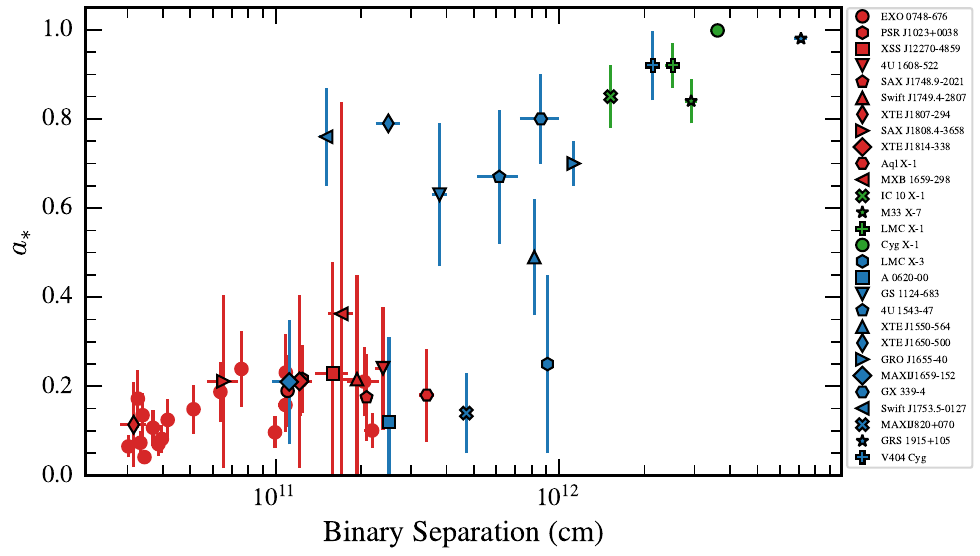}
\caption{The relation between the spin parameter $a_{*}$ and the orbital separation. The symbols are the same to \autoref{fig:porb}, except the red dots are the NS LMXBs without mass and/or radius measurements and with minimum companion mass.
\label{fig:sep}}
\end{figure*}

We found that the dimensionless spin parameter $a_{*}$ appears roughly correlated with the orbital period $P_{\rm orb}$ (\autoref{fig:porb}) including the NS and BH XRBs, which has been shown in \citet{fragos_origin_2015} with only the spin parameters of BH samples based on the continuum fitting method. Our samples include the LMXBs with fast rotating NSs, as well. The moment of inertia of an NS is dependent on the equation of state of ultra-dense matter. So far, the correct equation of state is not determined. For simplicity, the NS is assumed as a solid sphere, then its moment of inertia is given by $I = (2/5)M_\mathrm{NS}R_\mathrm{NS}^{2}$, where the $M_\mathrm{NS}$ and $R_\mathrm{NS}$ are the mass and radius of a NS. Then we got the dimensionless spin parameter of a NS $a_{*}=cJ/GM_\mathrm{NS}^{2}$, where $J=2\pi I\nu_{\rm spin}$ is the NS spin AM, and $\nu_{\rm spin}$ is the spin frequency. For LMXB without NS mass and/or radius measurement, we assume that the NS mass and radius are drawn from the distributions of current known samples. The NS mass are randomly sampled 1000 times from the double Gaussian distribution ($r_{1}*Norm(\mu_{1},\sigma_{1})+(1-r_{1})*Norm(\mu_{2},\sigma_{2})$) in \citet{rocha_max_2021} with $\mu_{1}=1.365$, $\sigma_{1}=0.109$, $r_{1}=0.498$ and $\mu_{2}=1.787$, $\sigma_{2}=0.314$, and the single Gaussian distribution in \citet{ozel_dense_2016} with $\mu=10.5$, $\sigma=1.5$ is used for NS radius sampling. Then the mean value and standard deviation of dimensionless spin parameters of those NSs can be calculated by using the mass and radius samples. In order to perform Spearman correlation rank test by taking into account the uncertainties of the two quantities, we used the Monte Carlo based methods in \citet{curran_monte_2014}. We resampled the data 5000 times, by randomly drawing from the Gaussian distribution, the center and sigma of which is taken as the measured value and the uncertainty. For the asymmetric uncertainties, we only use the superscript one. The Spearman coefficient between $a_{*}$ and $P_\mathrm{orb}$ is 0.64$\pm$0.07 at a significance of $4.83\pm0.68\sigma$, which demonstrate that there is a positive relation
between $a_{*}$ and $P_{\rm orb}$. The Spearman coefficients are 0.36$\pm$0.07 and 0.70$\pm$0.11 at a significance of $1.89\pm0.85\sigma$ and 3.14$\pm$0.69 for NS and BH samples, respectively. The positive relation is significant by combining the different types of compact objects. This result indicates that the spin parameter of the compact star somehow connects with the binary orbit. This relation seems to be flat above the orbital period of about 10 days due to the upper limit of permitted spin parameter. However, there is only one source with orbital period longer than 10 days (GRS 1915+105). 

The orbits of current XRBs are approximately circular, then the orbital separation is expressed as $2.9\times10^{11}M_{1}^{1/3}(1+q)^{1/3}P_\mathrm{orb,day}^{2/3}$ cm \citep{frank_accretion_2002}. The companion masses of most NS LMXBs are given as minimum masses in the literature for an assumed NS mass of 1.4 $M_{\odot}$ and inclination angle of 90$^{\circ}$ (see \autoref{tab:ns_par}). Here we randomly sample the NS mass from the double Gaussian distribution in \citet{rocha_max_2021} and the inclination angle in the range of 10$^{\circ}$--85$^{\circ}$, and use the mean value and standard deviation of the sample as the companion mass for further calculation. We further plotted the relation between the spin $a_{*}$ and the orbital separation (\autoref{fig:sep}). They show a positive relation as well with the Spearman coefficient $0.68\pm0.07$ at a significance of 5.22$\pm$0.73 $\sigma$, which is also obtained by using the Monte Carlo based method in \citet{curran_monte_2014}. The Spearman coefficients are $0.37\pm0.16$ and $0.70\pm0.11$ at a significance of $1.92\pm0.84\sigma$ and $3.14\pm0.67\sigma$ for NS and BH samples, respectively, which demonstrates the positive relation is weaker than that in the whole sample.

\begin{figure*}
\includegraphics[width=\linewidth]{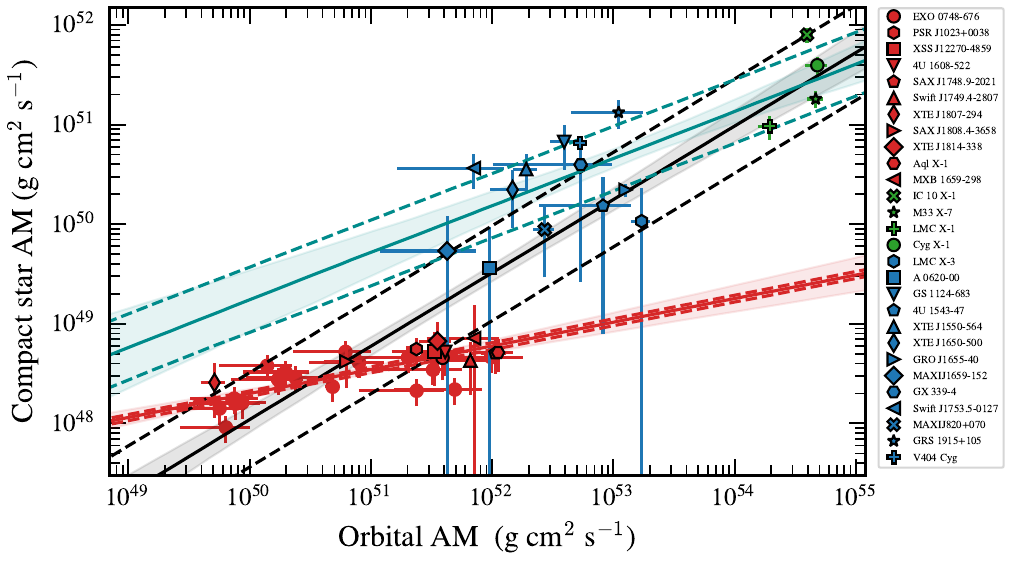}
\caption{The relation between the AM of the compact star and the binary orbit. The symbols are the same to \autoref{fig:sep}. The solid black line is the best-fitting linear model, the shadow region represents the uncertainties of the best-fitting parameters. The dashed line represents the best-fitting scatter of the data. The red and dark cyan lines represent the best-fitting results of NS and BH samples.
\label{fig:AM}}
\end{figure*}

The spin AM of the compact star ($J_{\rm C}$) and the orbital AM ($J_{\rm orb}$) of the binary system are expressed as,
\begin{align}
    \label{eq:AM}
    J_\mathrm{BH}&=a_{*}GM_\mathrm{BH}^{2}/c, \\
    J_\mathrm{NS}&=\frac{4}{5}\pi\nu_\mathrm{spin} M_\mathrm{NS}R_\mathrm{NS}^{2}, \\
    J_\mathrm{orb}&=(2\pi)^{-1/3}G^{2/3}M_{1}^{5/3}q(1+q)^{-1/3}P_\mathrm{orb}^{1/3}.
\end{align}
We then plotted the relation between $J_{\rm C}$ and $J_{\rm orb}$ (\autoref{fig:AM}). The Spearman coefficient is 0.73$\pm$0.10,  at a significance of $5.87\pm1.17 \sigma$,  which demonstrates that a positive relation between the spin and orbital AM exists by combining NS LMXBs, RLOF BH and wind-fed BH XRBs. The positive relations in NS and BH samples are weaker with the Spearman coefficient $0.47\pm0.15$ and $0.64\pm0.09$ at a significance of $2.53\pm0.87 \sigma$ and $2.74\pm0.52 \sigma$, respectively.  If we use a linear function in logarithmic scale, the best-fitting result is $\log{J_\mathrm{C}}=0.74^{+0.06}_{-0.06}\times\log{J_\mathrm{orb}}+11.26^{+2.98}_{-3.01}$, which is obtained by using a Bayesian inference package \texttt{UltraNest} \citep{buchner_ultranest_2021}. However, it seems that the NS LMXBs follow a shallower slope than BH XRBs. We then fitted the NS and BH samples separately, and the best-fitting functions are $\log{J_\mathrm{C}}=0.24^{+0.03}_{-0.04}\times\log{J_\mathrm{orb}}+36.27^{+2.08}_{-1.69}$ and $\log{J_\mathrm{C}}=0.47^{+0.09}_{-0.10}\times\log{J_\mathrm{orb}}+25.61^{+5.12}_{-4.86}$.

\section{DISCUSSION}
\begin{figure*}
\includegraphics[width=\linewidth]{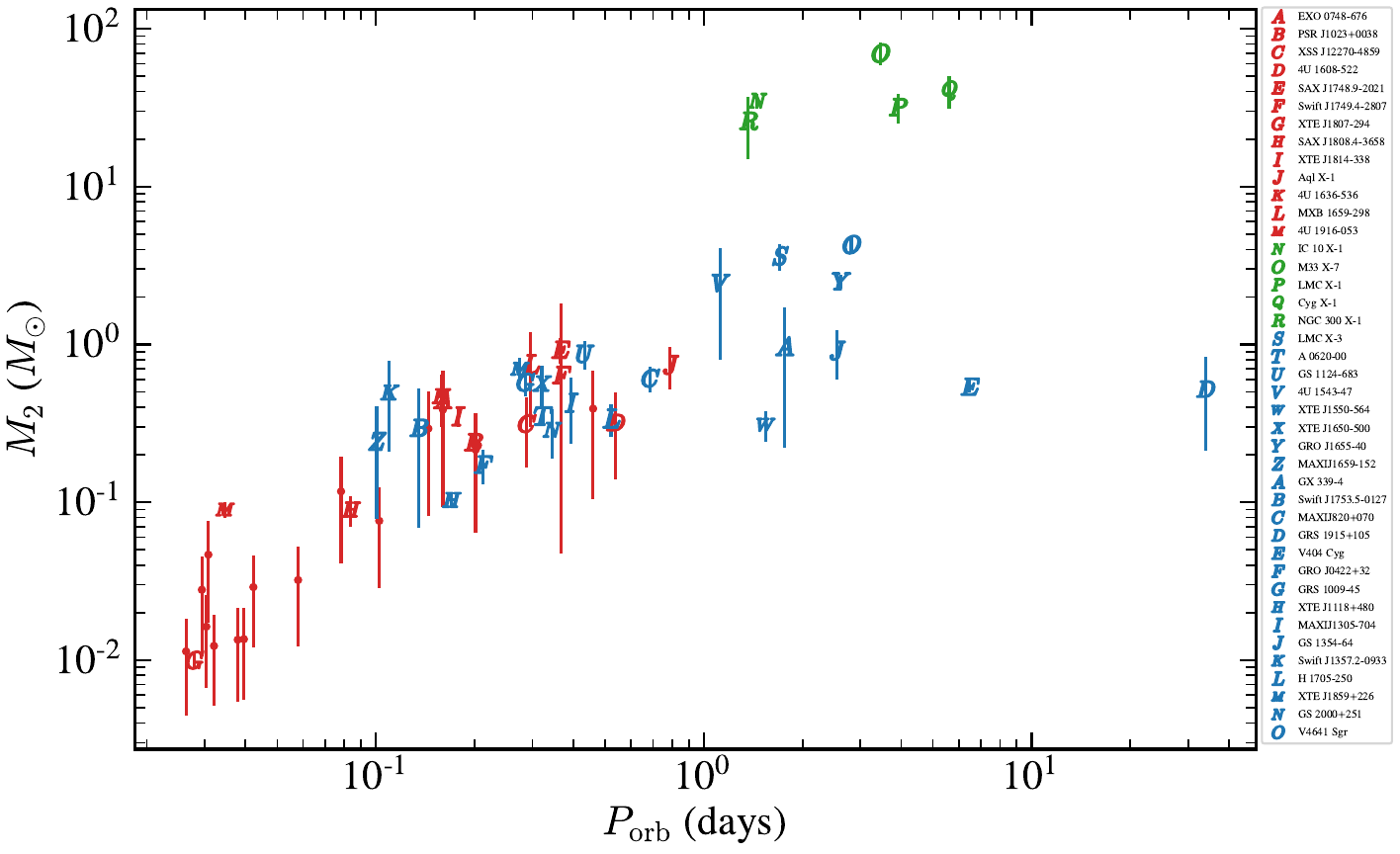}
\caption{The relation between companion mass and orbital period. The NS, wind-fed BH and RLOF BH systems are marked by red, green and blue, respectively. The red dots are the NS LMXBs with minimum companion mass measurements.
\label{fig:m2_p}}
\end{figure*}

\begin{figure*}
\includegraphics[width=\linewidth]{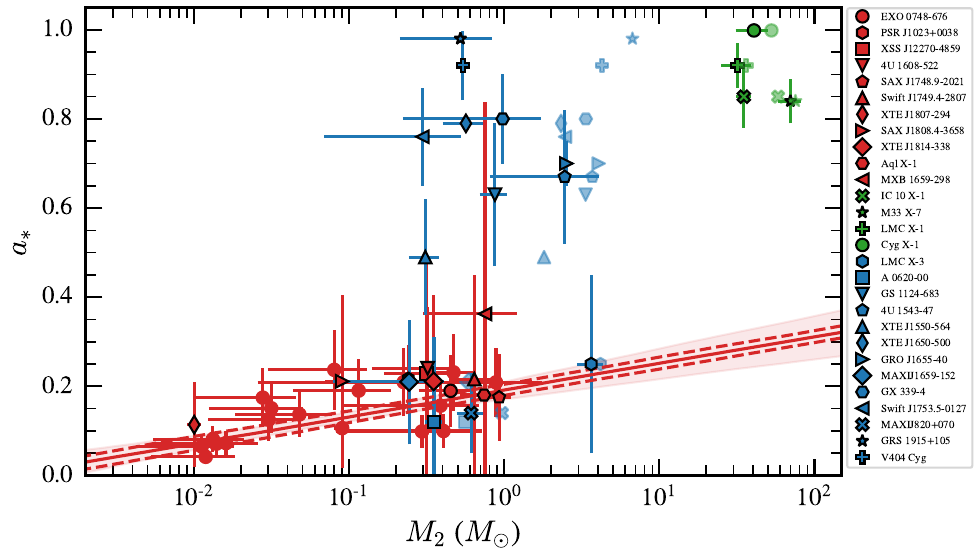}
\caption{The relation between the spin parameter of the compact star and the companion mass. The symbols are the same to \autoref{fig:sep}. The transparent green and blue markers represent the initial companion mass we estimated according to \citet{thorne_disk-accretion_1974}. The red line is the best-fitting linear function for NS samples.
\label{fig:m2_s}}
\end{figure*}

\begin{figure*}
\includegraphics[width=\linewidth]{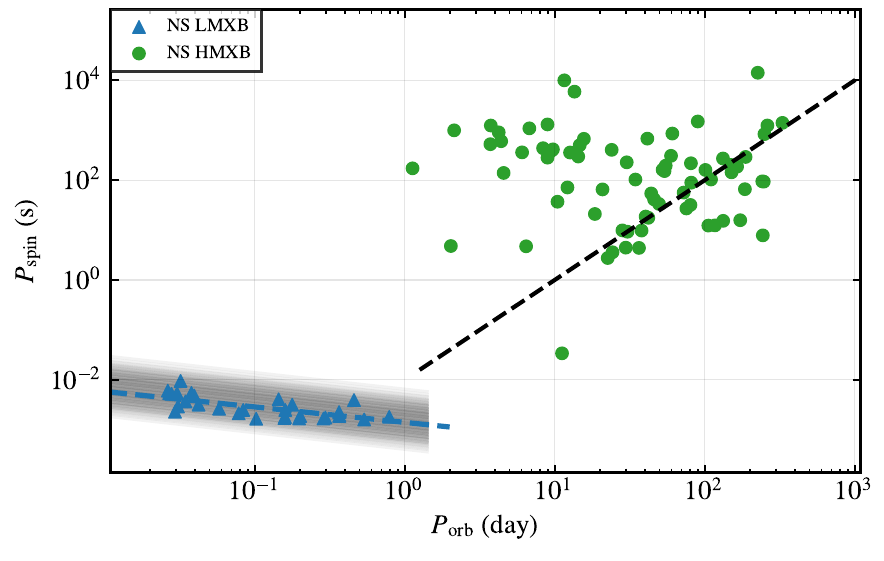}
\caption{The Corbet diagram including our NS LMXB samples. The dashed black line represents the empirical function $P_\mathrm{spin}=0.01P_\mathrm{orb}^{2}$ derived by \citet{corbet_1985}. The dashed blue line is the best-fitting result of NS LMXBs. The gray lines are the $P_\mathrm{spin}$ -- $P_\mathrm{orb}$ functions based on the assumption that the NS AM is proportional to the orbital AM. 
\label{fig:corbet}}
\end{figure*}

It is generally believed that the BH in an XRB is formed by the core collapse of a massive star, probably associated with an explosion of supernova or gamma-ray burst \citep[e.g.][]{wilson_stellar_1986,woosley_gamma-ray_1993}. Even the evolutionary pathways are different, the natal spin of the BH is mainly determined by AM of the stellar core if the AM is conserved during the collapse phase.

Asteroseismic measurements of single stars have shown that the stellar core rotates very slowly \citep[e.g.][]{beck_fast_2012,van_reeth_sensitivity_2018}, which indicates an efficient mechanism of AM transport must be at work to transport the AM from the stellar core to the envelope. The Tayler-Spruit magnetic dynamo \citep{spruit_dynamo_2002} is proposed to be the mechanism to slow down the core. According to the stellar evolution models including the Tayler-Spruit magnetic dynamo, the stellar core rotates slowly and then the spin parameter of the new born BH is very low \citep[e.g.][]{fuller_slowing_2019,fuller_most_2019,ma_angular_2019}.

In a binary system, the stellar core can be spun up by the tidal torque and becomes synchronized with the orbit period \citep[e.g. ][]{lee_discovery_2002,brown_gamma-ray_2007, axelsson_origin_2011}. However, the tidal spinning up is inefficient in the binary systems with orbital period longer than $\sim 1$ day \citep[e.g.][]{qin_spin_2018,fuller_spins_2022}. So the fast rotating stellar core only exists in a compact binary system. The mass of the companion star must be small to fit the short orbital period. The final spin parameter determined by the tidal locking correlates with the pre-explosion core mass and anti-correlates with the orbital period \citep{fuller_spins_2022}, which is inconsistent with the positive relation between the spin parameter and the orbital period (\autoref{fig:porb}).

In some supernova, the ejecta carries high specific AM and falls back onto the new born compact object to increase its AM \citep[e.g.][]{batta_formation_2017,chan_impact_2020}. However, the larger BH spin is expected in the shorter orbital separation \citep{schroder_black_2018}, which is contradictory to the observed results (see \autoref{fig:sep}). 

Overall, the natal spin of the BH in an XRB should be very low. This is consistent with the likely low spin parameters of the premerger BHs of the binary BHs detected by the LIGO/Virgo \citep{abbott_binary_2019,miller_low_2020}. The high spin must form after the BH birth. The accretion from the companion star is one of the most promising way \citep[e.g.][]{thorne_disk-accretion_1974}. However, it is generally believed the BH XRB with initially low-mass companion star cannot supply the required mass, and life time of a high-mass companion star is not long enough to significantly speed up the BH \citep[e.g. ][]{king_evolution_1999}. Then some scenarios have been proposed for this conundrum. For example, a hypercritical accretion ($\dot{M} \gg \dot{M}_{\rm Edd}$) phase occurs after the BH formed, and the initial companion mass is probably large ($>5M_{\odot}$) in some RLOF BH systems. In such cases, the lifetime and mass of the companions are both enough to spin up the BH to extreme value \citep{podsiadlowski_formation_2003,moreno_mendez_need_2011,fragos_origin_2015}. On the other hand, it is well known that the fast-rotating NS in a LMXB is spun up by the long-stable accretion \citep{bhattacharya_formation_1991}. 

Therefore the spin of the compact objects in different type XRBs are thought to relate with accretion process. In this work, we present some apparent positive relations between the spin parameter $a_{*}$ and the orbital parameters (such as orbital period and orbital separation), and between spin AM and orbital AM by combining the NS and BH XRBs. We notice that these positive correlations of the total sample are more significant than the subclass, which indicates different detailed mechanism works in different types of XRBs. The bimodal distribution of BH spin parameters also indicates that the low and high spin BHs experienced different evolution and accretion history. We then explore the spin evolution of NS and BH in XRBs from these relations.

\subsection{NS LMXB}
 The long and stable mass transfer through the RLOF in the NS LMXBs can provide sufficient mass to spin up the NS to rotation rates at hundreds of Hz \citep[see reviews in ][]{bhattacharya_formation_1991,patruno_accelerated_2012,campana_accreting_2018}. According to \autoref{eq:ns_spinup} of \citet{tauris_formation_2012}, only $\sim0.06M_{\odot}$ and $\sim$ 50 Myr are needed to speed up a NS with spin period about 3 ms at the 10\% Eddtington limit (a few times $10^{-9} M_{\odot} ~\rm yr^{-1}$):
\begin{align}
\label{eq:ns_spinup}
\Delta M &\simeq 0.06\left(\frac{M_\mathrm{NS}}{1.4M_{\odot}}\right)^{1/3}\left(\frac{P_\mathrm{spin}}{3ms}\right)^{-4/3} ~\rm M_{\odot}, \\
t &\simeq  50~B_{8}^{-8/7} \left(\frac{\dot{M}}{0.1\dot{M}_\mathrm{Edd}}\right)^{-3/7} \left(\frac{M_\mathrm{NS}}{1.4M_{\odot}}\right)^{17/7} \mathrm{Myr},
\end{align}
where $B_8$ is the magnetic field of the NS in $10^{8}~\rm G$, $\dot{M}$ is the mass accretion rate, and $P_{\rm spin}$ is the NS spin period. Therefore, NS in LMXB can be spun up to about several ms during the LMXB phase, which is suggested to explain the millisecond pulsar in binary after the discovery of the first one \citep{alpar_new_1982,radhakrishnan_origin_1982}. The discovery of the first accreting millisecond pulsar (AMSP) SAX J1808.4$-$3658 further supports this link \citep{wijnands_millisecond_1998,chakrabarty_two-hour_1998}.

Under the accretion spinning up scenario, the final spin frequency is determined by the equilibrium configuration when the angular velocity of the NS is equal to the Keplerian angular velocity of the accretion matter at the the magnetospheric boundary \citep{bhattacharya_formation_1991,tauris_formation_2012}:
\begin{equation}
\label{eq:ns_f}
    \nu_{eq} \simeq 714~ B_{8}^{-6/7}\left(\frac{\dot{M}}{0.1\dot{M}_\mathrm{Edd}}\right)^{3/7}\left(\frac{M_\mathrm{NS}}{1.4M_{\odot}}\right)^{5/7}R_{13}^{-18/7} ~\rm Hz
\end{equation}
where $R_{13}$ is the NS radius in units of 13 km. Equation~\ref{eq:ns_f} shows that the equilibrium spin frequency mainly depends on the mass accretion rate, since we expect similar magnetic field, NS mass and radius in NS LMXBs. The long term average mass accretion is expected to be proportional to the mass transfer rate from the companion star, which is believed to correlate with the companion mass and the orbital period \citep{king_evolution_1988}. The orbital period strongly depends on the companion mass (see \autoref{fig:m2_p}). The NS mass and radius is expected to be almost unchanged during the accretion spinning-up phase, so the AM of the NS gained due to accretion should correlate with the companion mass. We plotted the relation between the spin parameter $a_{*}$ and the companion mass $M_{2}$ (\autoref{fig:m2_s}). It seems that there is a weak positive correlation between $a_{*}$ and $M_{2}$ in NS sample, the Spearman coefficient is 0.34$\pm$0.16 at a significance of $1.77\pm0.88\sigma$, which supports the scenario that the current AM of NS in LMXB is related with mass of the companion. It seems that the relation saturate at around $a_{*}\sim 0.2$.  If we use a linear function to fit the NS sample, the best-fitting result is $a_{*}=0.06\pm0.02\log(M_{2})+0.19\pm0.03$. Since the $M_\mathrm{NS}$ varies little, the orbital AM mainly depends on $P_\mathrm{orb}$ and $M_{2}$. As a result, the NS AM should positively correlate with the orbital AM (\autoref{fig:AM}).

The Corbet diagram (spin period $P_\mathrm{spin}$  vs. orbital period $P_\mathrm{orb}$) has been used as a great tool for studying the formation,evolution and accretion history of NS HMXB \citep{corbet_1984}. We used the most updated HMXB catalog \citep{fortin2023} and our NS LMXB samples to build the Corbet diagram. Our NS LMXB sample obviously differs from the HMXB sample on the Corbet diagram (see \autoref{fig:corbet}). The $P_\mathrm{spin}$ and $P_\mathrm{orb}$ of NS LMXBs follow a negative correlation with a Spearman coefficient -0.66 at a significance of 3.90 $\sigma$, which is consistent with the positive relations we found between the spin and orbital parameters (\autoref{fig:porb} and \autoref{fig:sep}). We fitted the data of NS LMXBs with a function $\log(P_\mathrm{spin})=a\times\log(P_\mathrm{orb})+b$, and the best-fitting slope $a$ is $-0.31\pm$0.06. If the NS AM is simply assumed to be proportional to the orbital AM, then $P_\mathrm{spin}\propto P_\mathrm{orb}^{-1/3}q^{-1}(1+q)^{-1/3}M_\mathrm{NS}^{-2/3}R_\mathrm{NS}^{2}$, which is consistent with the best-fitting slope. We then added this function on the Corbet diagram with 0.01<q<0.1, 1.<$M_\mathrm{NS}$<2.3 and 9<$R_\mathrm{NS}$<12 (grey lines in \autoref{fig:corbet}), and the NS LMXBs follow this function quit well. The NS LMXBs are obviously opposite to the positive relation between $P_\mathrm{spin}$ and $P_\mathrm{orb}$ in HMXBs discovered by \citet{corbet_1984}, which demonstrates that the evolution and accretion history are different between NS LMXBs and HMXBs. However, as more and more HMXBs detected, the HMXBs show diverse distribution on the Corbet diagram \citep[][and also \autoref{fig:corbet}]{corbet_1986}. The accretion flow may play an important role in determining the $P_\mathrm{spin}$-$P_\mathrm{orb}$ relation of NS HMXB \citep[e.g.][]{cheng_2014}. So the obvious difference between LMXBs and HMXBs on the Corbet diagram may also indicates the different accretion processes.

\subsection{BH XRBs} 
In the BH XRB case, it is usually assumed that the matters falling onto the BH carry the AM when they are at the ISCO. Although different accretion flows show some differences during the spinning-up process \citep[e.g.][]{thorne_disk-accretion_1974,abramowicz_spin-up_1980,popham_advection-dominated_1998,sadowski_spinning_2011}, the final spin parameter of the BH mainly depends on the total accreted mass as a fraction of the initial BH mass, even for hypercritical accretion ($\gg \dot{M}_{\rm Edd}$). For example, for a 5 $M_{\odot}$ BH with zero natal spin, the required accreted mass will be 1 $M_{\odot}$ to achieve BH spin 0.5 \citep[see][]{thorne_disk-accretion_1974,belczynski_black_2008}. 

The RLOF BHs with low spin parameters ($<$0.3) seem to follow a similar $M_{2}$-$a_{*}$ relation to the NS LMXBs (\autoref{fig:m2_s}). Those RLOF BHs probably have an initial low-mass companion star which is not massive enough to feed the BH to achieve high spin parameter. If the current BH spin are mostly gained from the low level accretion ($< \dot{M}_{\rm Edd}$) after the BH formed, their ages of the XRB phase should be very short. Actually, previous studies of a few RLOF BHs with low spin do indicate short accretion time. For instance, A0620-00 \citep{gonzalez_hernandez_chemical_2004} and LMC X-3 \citep{sorensen_unraveling_2017} are thought to have short accretion time ($\sim 10^{7}$ yr ). So similar to the NS LMXBs, the spin value of RLOF BHs ($a_{*}<0.3$) also depends on the mass transfer rate, then the positive relations between the spin parameters and the orbital parameters are expected.

There are more RLOF BHs with large spin ($>0.5$; see, e.g., \autoref{fig:m2_s}). \citet{fragos_origin_2015} have argued that the current high BH spin parameters in some BH LMXBs are acquired through accretion \citep[see also][]{podsiadlowski_formation_2003}. For example, the BH in GRS 1915+105 could achieve the near-maximal spin by doubling its mass via accretion. A plausible scenario for BH LMXBs with high spin parameters is that, at the initial stage of the XRB phase, the system is required to be an intermediate-mass X-ray binary (IMXB) with a comparable companion mass to the BH ($q\sim1$), and then the companion star must lose most of its mass to feed the BH. It is expected that a BH IMXB experiences a period with hypercritical accretion rate ($\gg \dot{M}_\mathrm{Edd}$) due to the thermal timescale mass transfer ($\sim10^{-6}$--$10^{-5} M_{\odot}/yr$), when the BH accretes large amounts of mass during a short duration ($\sim$ a few Myr) \citep{podsiadlowski_evolutionary_2002,xu_thermal_2007}. Some ultraluminous X-ray sources (ULXs) in external galaxies are possibly such IMXBs, which are evolving towards to BH LMXBs \citep{king_evolutionary_2000,king_ultraluminous_2001,kalogera_observational_2004,begelman_nature_2006,king_black_2016}.

On the other hand, all the measured BH spin parameters in the known wind-fed BH XRBs are very high ($>0.8$; see \autoref{tab:bh_par}). Their companion stars are massive enough to spin up the BH to such high value. However, even through the Eddington accretion, the total accreted mass during the whole lifetime of the massive companion star is less than 1 $M_{\odot}$, which is not enough to spin up the BH significantly. So hypercritical accretion ($\gg \dot{M}_\mathrm{Edd}$) are needed to achieve the observed spin parameter \citep{moreno_mendez_need_2011,qin_hypercritical_2022}. It has been proposed that these wind-fed BH XRBs have experienced a hypercritical accretion phase, during which the BH can be spun up to extreme spin value in a short timescale \citep{moreno_mendez_how_2022}. Some extragalactic ULXs might be wind-fed BH XRB in this phase. 

Therefore, both high-spin RLOF and wind-fed BHs are expected to experience a short hypercritical accretion phase to gain enough AM. Even the detailed properties of the hypercritical accretion flow is unclear, the obtained AM by the BH is determined by the total accreted mass during this spinning-up phase. The amount of accreted mass is mainly determined by the mass transfer rate and timescale, which crucially depends on the nature of the companion star. The initial companion mass constrains the upper limit of the total accreted mass. So the spin parameter should relate with the initial companion mass. If the current AM of BH are gained from accretion, we can estimate the lower limit of the initial companion mass by adding up the accreted mass. 
Using Eq.2 of \citet{thorne_disk-accretion_1974}, we compute the accreted mass by assuming that the current AM of the BH is mostly acquired by accretion, and subsequently estimate the ``initial'' companion mass (transparent symbols in \autoref{fig:m2_s}). It seems that the current BH spin positively relates with estimated ``initial'' companion mass. However, they do not follow the $a_{*}$-$M_{2}$ relation of NS samples, which indicates that the evolution and accretion history are indeed different.

For the conservative mass transfer, the change of the orbital separation can be roughly expressed as \citep[see also Eq. 4.14 in][]{frank_accretion_2002},
\begin{equation}
     \frac{\Delta a}{a}=\frac{\Delta M_{2}}{M_{2}} \left (1-\frac{M_{2}}{M_{1}} \right ).
 \end{equation}
For high spin RLOF BHs, the
masses of the companion star and BH are comparable in the IMXB during the short-lived hypercritical accretion phase, so the orbital separation remains almost constant during this short period. Since then, the loss mass of the companion star is neglectable. It is therefore expected to observe the positive relationship between the spin parameter and orbital separation/orbital period, which relates with the initial companion mass. For wind-fed BH systems, since the fractional mass loss of the companion star is very small for wind-fed BH systems (see \autoref{fig:m2_s}), the orbital separation should change insignificantly. Although the above estimation is crude, the main conclusion should not change even with more thorough analysis in which the detailed accretion history and evolution during the short-lived hypercritical accretion phase is taken into account for both RLOF and wind-fed BH systems. In a word, the positive relation between the spin parameter and orbital parameters is plausible under the accretion spinning-up scenario. However, the low and high spin BH XRBs have different evolution paths and accretion history, which is consistent with the bimodal distribution of BH spin parameters (\autoref{fig:hist}).

Generally speaking, binary system with longer orbital period harbors more massive companion star (\autoref{fig:m2_p}). However, we notice that there two BH XRBs (V404 Cyg and GRS 1915+105) apparently deviate from the $P_\mathrm{orb}$ -- $M_{2}$ relation. These two sources have longer $P_\mathrm{orb}$ than others in our sample, and are the only two RLOF BHs with spin parameters larger than 0.9. They may experience different accretion history and/or evolution pathway after the hypercritical accretion phase. On the other hand, it is possible that the relation between $M_{2}$ and $P_\mathrm{orb}$ shows two distinct branches (similar to ``$\Lambda$'' shape) below and above $\sim$ 2 days in RLOF BH XRBs (\autoref{fig:m2_p}). This pivot period probably corresponds to the bifurcation period $P_\mathrm{bif}$ \citep[$\sim$1--2 days,][]{ma_bifurcation_2009}, which is critical in determining the evolution pathway of the BH XRBs \citep[e.g.][]{fragos_origin_2015}. The two sources (V404 Cyg and GRS 1915+105) apparently at the right branch, which indicates that they share a similar evolution pathway.

\section{Summary and Implication}
We collect measurements of the spin parameters of BH XRBs and the spin frequencies of fast rotating NS LMXBs ($\nu_\mathrm{spin}>100$ Hz) from the literature, and investigate relations between the spin parameters and the orbital parameters of the sample. We confirm that there is a positive relation between the spin parameter $a_{*}$ and orbital period $P_\mathrm{orb}$ \citep{fragos_origin_2015} by including different XRB systems (\autoref{fig:porb}), irrespective which method is used to measure the BH spin parameter. We also find a positive relation between the spin parameter and the orbital separation (\autoref{fig:sep}) in the sample. The AM of the compact star shows an apparent positive relation with the orbital AM by combining the NS LMXBs, the RLOF and wind-fed BH XRBs (\autoref{fig:AM}). These positive relations imply that the formation of the current spin of the different compact objects may all relate with accretion process, even though the detailed mechanisms are still uncertain. 

Current theory supports that the stellar core before collapsing into BH rotates very slowly \citep[e.g.][]{fuller_slowing_2019}. Other mechanisms such as tidal spinning up or fall-back accretion also contradict with observations \citep[e.g.][]{moreno_mendez_how_2022}. Then the new born BH is expected to have a very low spin parameter. The current BH spin is formed after the BH birth. Accretion is the promising mechanism for spinning up the BH. On the other hand, the long stable accretion process is well-known for accounting for the fast rotating NS in LMXB \citep[e.g.][]{bhattacharya_formation_1991}. The positive relations we found between spin parameters and orbital parameters (\autoref{fig:porb},\autoref{fig:sep} and \autoref{fig:AM}) are plausible to form under the accretion spinning-up framework.

We infer that the low spin BH ($a_{*}<0.3$) is spun up by the low-level accretion ($< \dot{M}_\mathrm{Edd}$) in RLOF BH XRBs with an initial low-mass companion, which is similar to the NS LMXB. The current high BH spin parameter ($a_{*}>0.5$) seen in RLOF and wind-fed BH XRBs can be achieved by a short-lived hypercritical accretion ($\gg \dot{M}_\mathrm{Edd}$) with an initial intermediate/high-mass companion. The above scenario is consistent with the bimodal distribution of BH spin parameters. Some extragalactic ULXs may correspond to such hypercritical accretion phases, a census of BH spin parameters in those ULXs would be crucial in testing the BH spin evolution both for RLOF and wind-fed BH XRBs with high spin parameters. However, there is no BH spin measurement of ULX so far, since the sensitivity of current observation is not good enough. Current X-ray mission such as $Insight$-HXMT, NICER are capable to constrain the BH spin for a Galactic BH ULX. It thus helpful to observe the emergence of a Galactic BH in the ULX phase in the near future and to measure BH spin with X-ray observation.

\section*{Acknowledgements}
This work was support in part of the Natural Science Foundation of China (grants U1838203, and U1938114). Z. Y. was also supported by the Youth Innovation Promotion Association of Chinese Academy of Sciences and funds for key programs of Shanghai astronomical observatory. WZ acknowledges the support by the Strategic Pioneer Program on Space Science, Chinese Academy of Sciences through grant XDA15052100.

\section*{Data Availability}
The data used in this work are collected from literature and listed in \autoref{tab:bh_par} and \autoref{tab:ns_par}.








\bsp	
\label{lastpage}
\end{document}